\documentclass{aa}
\usepackage{graphics}

\def\masl{\ifmmode  {\rm M_{\sun}yr^{-1}} \else ${\rm M_{\sun}yr^{-1}}$\fi}

\def\mdot{\ifmmode  \dot{M} \else $\dot{M}$\fi}
\def\msun{\ifmmode {\rm M}_{\odot} \else M$_{\odot}$\fi}
\def\vinf{\ifmmode v_{\infty} \else $v_{\infty}$\fi}
\def\teff{\ifmmode T_{\rm eff} \else $T_{\rm eff}$\fi}
\def\logg{\ifmmode \log g \else $\log g$\fi}
\def\loggeff{\ifmmode \log g_{\rm eff} \else $\log g_{\rm eff}$\fi}
\def\rstar{\ifmmode R_{\star} \else $R_{\star}$\fi}
\def\tstar{\ifmmode T_{\star} \else $T_{\star}$\fi}
\def\lstar{\ifmmode L_{\star} \else $L_{\star}$\fi}
\def\mstar{\ifmmode M_{\star} \else $M_{\star}$\fi}
\def\rsun{\ifmmode {\rm R}_{\odot} \else ${\rm R}_{\odot}$\fi}
\def\lsun{\ifmmode {\rm L}_{\odot} \else ${\rm L}_{\odot}$\fi}
\def\kms{\ifmmode {\rm km\,s^{-1}} \else $\rm km\,s^{-1}$\fi}
\def\hei{He\,{\sc i}}
\def\heii{He\,{\sc ii}}
\def\cii{C\,{\sc ii}}
\def\ciii{C\,{\sc iii}}
\def\civ{C\,{\sc iv}}
\def\oiii{O\,{\sc iii}}

\def\cav{Ca\,{\sc v}}
\def\fevi{Fe\,{\sc vi}}
\def\niv{Ni\,{\sc vi}}
\def\cmf{{\sc cmfgen}}
\def\isa{{\sc isa-wind}}
\def\la{$\lambda$}

\begin{document}
\thesaurus{06 (08.01.3; 08.02.4; 08.06.3; 08.09.2 $\gamma$~Velorum ;08.13.2; 08.23.2)}

\title{The $\gamma$~Velorum binary system
\thanks{Based on observations collected at the
European Southern Observatory at La Silla, Chile. ESO proposals
Nrs. 56.D-327, 57.D-517 and 56.D-0700}
}
\subtitle{II. WR stellar parameters and the photon loss mechanism} 
\author{Orsola De Marco$^{1,2}$, W.~Schmutz$^{3,2}$, 
P.A. Crowther$^1$, D.J. Hillier$^4$, L.~Dessart$^1$\thanks{Present address:
D\'epartement de Physique, Universit\'e Laval and Observatoire du 
Mont M\'egantic, Quebec, QC G1K 7P4, Canada.},
A. de~Koter$^5$ \and J. Schweickhardt$^6$}
\institute{
$^1$Department of Physics and Astronomy, University College London, 
    Gower Street, London WC1E 6BT, UK \\
$^2$Institut f\"ur Astronomie, ETH-Zentrum, Scheuchzerstrasse 7, CH-8092 Z\"urich, Switzerland \\
$^3$PMOD/WRC, Dorfstrasse 33, Davos Dorf CH-7260, Switzerland\\
$^4$Department of Physics and Astronomy, University of Pittsburgh, 
    3941 O'Hara Street, Pittsburgh PA 15260, USA \\
$^5$Astronomical Institute Anton Pannekoek, University of Amsterdam, 
    Kruislaan 403, NL-1098 SJ, Amsterdam, The Netherlands \\
$^6$Landessternwarte Heidelberg-K\"onigstuhl, D-69117
    Heidelberg, Germany }
\mail{od@star.ucl.ac.uk}
\date{Received  25 January 2000 / Accepted }
\titlerunning{The $\gamma$~Velorum binary system. II}
\authorrunning{O. De Marco et. al.}
\maketitle


\begin{abstract}
In this paper we derive stellar parameters for the Wolf-Rayet star in the
$\gamma$~Velorum binary system (WR11), from a detailed non-LTE model of its
optical and infrared spectra. Compared to the study of Schaerer et al.,
the parameters of the WC8 star are revised to a hotter effective temperature
($T_{\rm eff} \sim 57$~kK),
a higher luminosity ($\log(L$/L$_{\odot})$ = 5.00), and a 
lower mass loss rate (log($\dot{M}$ / M$_{\odot}$/yr) = --5.0 - using a 10\% 
clumping filling factor. These changes lead to a significant decrease
         in wind efficiency number, from 144 to 7, 
         so that the driving mechanism of the wind 
         of this WR star may be simply radiation pressure on lines. 
The derived spectroscopic luminosity
is found to be 40\% lower than that derived by De Marco \& Schmutz through the
mass-luminosity relationship for WR stars ($\log(L$/L$_{\odot})$ = 5.2).

The paper furthermore presents a comparison of the independently-developed
modelling programs, \cmf\ and \isa . Overall, there seems 
to be very reasonable agreement between the derived parameters for WR11,
except for the carbon content, which is 2 times higher for \cmf\ 
(C/He=0.15 vs. 0.06, by number).
The comparison also confirms a disparity in the predicted 
flux at $\lambda$$<$400 \AA , found by Crowther et al., which will have 
effects on several nebular line strengths. 

The paper also presents the first independent check of the photon
loss mechanism proposed by Schmutz. We conclude that, not only is it
important to include very many lines to realistically model line blanketing,
but in particular those ones that critically interact with strong resonance
lines (e.g. \heii\ \la 303.78). The inclusion of these latter lines
may significantly alter the wind ionization structure.

\end{abstract}


\section{Introduction}\label{sec:intro}

The double-lined spectroscopic binary $\gamma$~Velorum (WC8 + O7.5) can be used 
to derive the mass of the Wolf-Rayet star, a quantity which cannot be 
directly determined
in the case of single objects, but which provides an important constraint 
in computations of the late stages of massive star evolution.
From it, and the mass-luminosity relationship for WR 
stars (Schaerer \& Maeder 1992) a value for the WR luminosity can be
derived and compared to that derived from spectroscopic analyses.

Orbital parameters together with the mass ratio were 
determined by Schmutz et al. (1997),
while the 7.3$\sigma$ {\sc hipparcos} distance allowed Schaerer et al. (1997) 
to derive the system's total mass ($\cal M$(O+WR)=29.5$\pm$15.9~\msun ),
the absolute visual magnitude ($M_V$(O+WR)=$-5.39$) and to carry out
a preliminary spectroscopic analysis of both components. 
Schaerer et al. (1997), however, modelled only
the WR \hei\ \la 4471 / \heii\ \la 4541 line strength ratio,
{\it assuming} the mass-loss rate and the
carbon abundance. Additionally, line blanketing was not 
accounted for by their model
and their calculation used simple helium and carbon model atoms.
Their stellar parameter determination should therefore be considered 
preliminary.

De Marco \& Schmutz (1999; hereafter Paper~I) carried out a
full spectroscopic analysis of the O star through which its 
parameters were determined {\it simultaneously}
with the light ratio between the WR and O stellar components.
From a solution of the hydrodynamic wind equations, the mass of the O
star was also derived and found to be consistent with the evolutionary
O star mass. From it and the mass ratio of
Schmutz et al. (1997), 
the mass of the WR star could be calculated (M=9.0$\pm$1.0~M$_\odot$).
The combined mass of the system was therefore derived to be 39~M$_\odot$,
higher than the value of Schaerer et al. (1997), but consistent within
their uncertainty. From the mass of the WR star and
the theoretical mass-luminosity relation of
Schaerer \& Maeder (1992), a value for the WR bolometric luminosity was
derived (log($L$/L$_\odot$) = 5.2), significantly
higher than
the spectroscopic luminosity derived by Schaerer et al. (1997; log($L$/L$_\odot$) = 4.8).
Finally, using the synthetic O star spectrum, 
the WR11 spectrum was de-convolved from the O star component 
allowing the WR emission lines to be measured with greater accuracy.

In this paper we determine the stellar
parameters of the WR star using the clumped, line-blanketed stellar
atmosphere model \cmf\ (Hillier 1987, Hillier \& Miller 1998). 
Ultimately, higher  spectroscopic luminosities 
would lower the wind performance number, indicating
that a smaller number of photon scatterings is needed to
achieve the radiative driving of the wind. 
This would facilitate the derivation of the wind velocity law.
$\gamma$~Velorum provides us with a new tool to test the model luminosity 
with an independent method in such a way as to impose a further
check on the wind momentun number. 

A second aim of this paper is to continue the comparison of the 
non-LTE stellar atmosphere code \cmf\ and the independently developed
Sobolev approximation code \isa\ (de~Koter et al. 1993, 1997), which is
used in combination with the line-blanketing Monte Carlo code of 
Schmutz (1991). This study follows from the comparison
carried out by Crowther et al. (1999) for the WN8 star WR124.

Finally, 
Schmutz (1997) showed that line-line interaction between \heii\ Ly$\alpha$
and metal lines at similar wavelengths can lead to a larger spectral luminosity.
The inclusion of this interaction in the models, an effect that he called ``photon loss",
showed that WR6 might have a larger luminosity than previously thought, and 
allowed him
to calculate a velocity law for its outer wind.
Since line-line interaction is included in the \cmf\ model code, we can test whether the
approximation of Schmutz (1997) is valid and quantify the importance
of the photon loss effect in the case of WR11.

In Sect.~\ref{sec:obs} we summarise the observations, while in Sect.~\ref{sec:red}
we discuss the light ratio between the O and WR stars and the reddening
towards the system. 
In Sect.~\ref{sec:mod} we describe the model atmosphere codes used in the 
determination of the stellar parameters. We present our results in
Sect.~\ref{sec:res}, together with the code comparison, 
while in Sect.~\ref{sec:pholos} the effect of 
photon loss is studied. We finally
draw our conclusions in Sect.~\ref{sec:conc}. 


\section{Observations and reduction}\label{sec:obs}
Our optical observations of $\gamma$~Velorum 
are the same as those used by Schmutz et al. (1997), where a full account
of the data reduction can be found. In summary,
several spectra have been obtained at the 50~cm ESO telescope in conjunction
with the Heidelberg Extended Range Optical Spectrograph ({\sc heros})
in the
ranges 3500-5500 \AA\ and 
5800-8600 \AA\ covering the binary period. 
The resolution is R=20\,000, 
while the S/N ratio ranges between 100 at 3600 \AA\ and 250 at 6000 \AA . 
The phase-averaged spectrum of the $\gamma$~Velorum system was rectified as described in
Paper~I. It is important to remind that rectification of WC stars is a very
delicate operation because of blending of large numbers of broad lines, which
leave few identifiable continuum points. Moreover, spectral ranges at the end of
instrumental orders (5300-5500~\AA\ and 5800-5900~\AA ), suffer from additional uncertainty.
Unfortunately this affects the diagnostic lines \hei\ \la 5876, \heii\ \la 5412,
and \civ\ \la 5471. Extreme care was taken in rectifying these
ranges and consistency between these lines and those residing at other
wavelengths was insured.
 
Observations of the 10830~\AA\ \hei\ line were obtained at the European
Southern Observatory (ESO) New Technology Telescope on January 12 and 13, 1996,
with the {\sc emmi} instrument. {\sc emmi} was used in the {\sc remd} mode
with grating No. 7, a slit-width of 1.5$^{''}$
and the spectrum recorded on ESO CCD No. 36. 
Two exposures, each lasting 10~s, were taken. Wavelength 
calibration was obtained with respect to a ThAr arc lamp, to
achieve a resolution of 2~\AA . Relative flux calibration was obtained with 
respect to a 180~s exposure of the standard star $\theta$~Crt. 

Mid-IR data of WR11 were obtained as part 
of Guest Observer programme WRSTARS (P.I. van der Hucht),
with the Short Wavelength Spectrograph (SWS; de Graauw et al. 1996)
on board 
the ESA Infrared Space Observatory (ISO; Kessler et al. 1996).
The SWS AOT6 observing mode was
used to achieve full grating resolution, $\lambda/\Delta\lambda \;
\sim \; 1300-2500$, with the wavelength coverage 2.38--45.0$\mu$m.
The observations were flux calibrated to an accuracy of 5\% and $\sim$20\%
for the low and high wavelength ends, respectively. 
The possibility that the nearby $\gamma^1$~Velorum might have contaminated the
ISO observations of $\gamma$~Velorum was investigated, but with an aperture of
20$^{\prime \prime}$ centered on $\gamma$~Velorum it is not possible
that light from 
$\gamma^1$~Velorum, at 40$^{\prime \prime}$
distance, could have entered the aperture. 
For full details on the observations and data reduction procedures, we 
refer the reader to Morris et al. (2000).

Finally, we have obtained a flux-calibrated UV dataset for WR11 as follows.
WR11 was too bright for absolutely flux calibrated observations with IUE/LORES.
Consequently we have matched IUE observations from July 1988, to low resolution 
S2-68 observations from 1973 at phase 0.3 (Willis \& Wilson 1976). 


\section{Light ratio correction and reddening}\label{sec:red}

In this section we discuss the O-WR light ratio of $\gamma$~Velorum. We also re-evaluate
the interstellar reddening that was adopted in Paper~I, namely
$E(b-v)=0.03$ mag, obtained from assumed intrinsic colours 
(van der Hucht et al. 1988).

\begin{figure}
\vspace{8cm}
\includegraphics{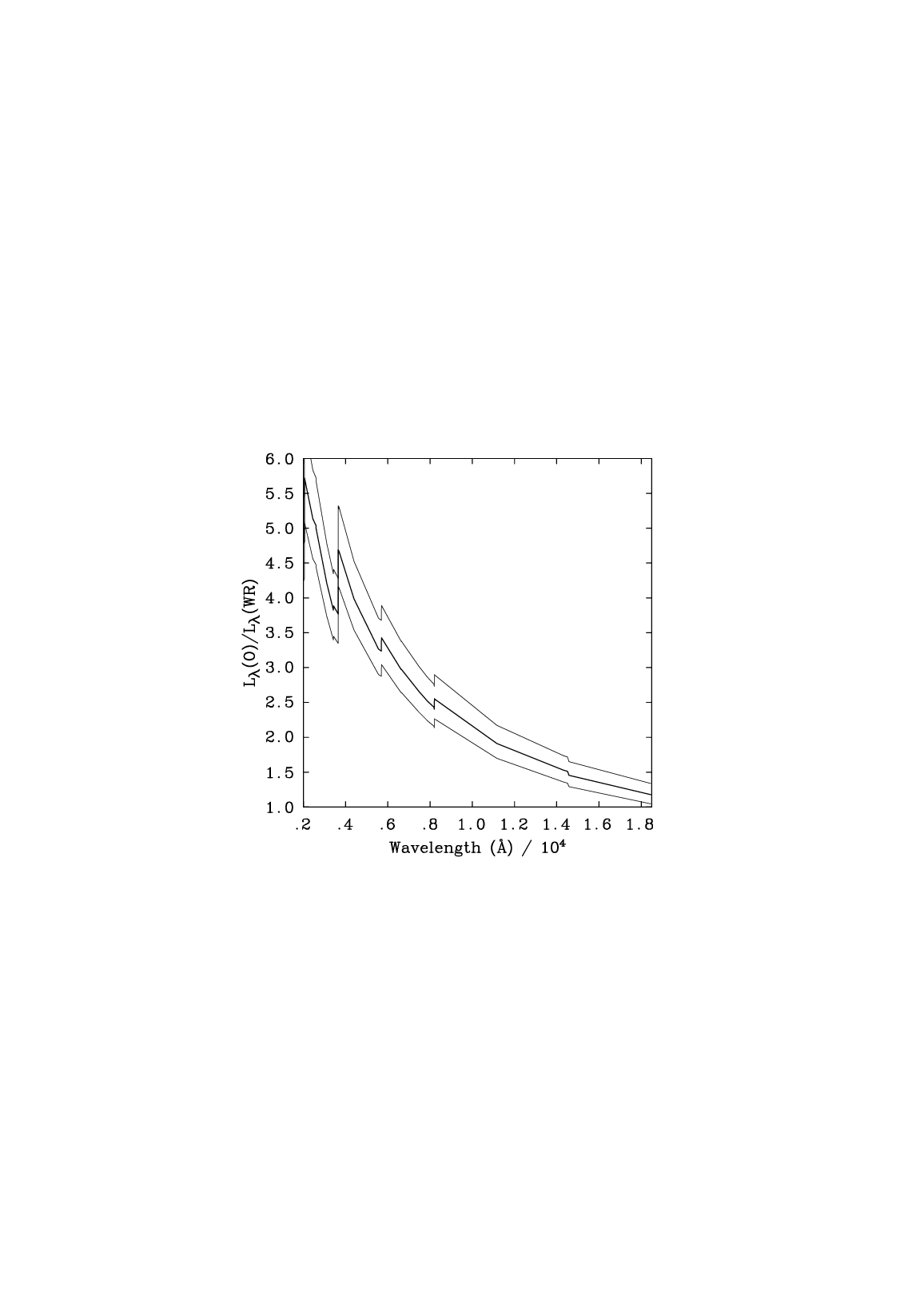}
\caption{The light ratio, $L_\lambda$(O)/$L_\lambda$(WR), as a function of
wavelength. The thinner lines show the error bars derived
from the original error limits of Paper~I}
\label{light_ratio}
\end{figure}

\begin{figure}
\vspace{6cm}
\includegraphics{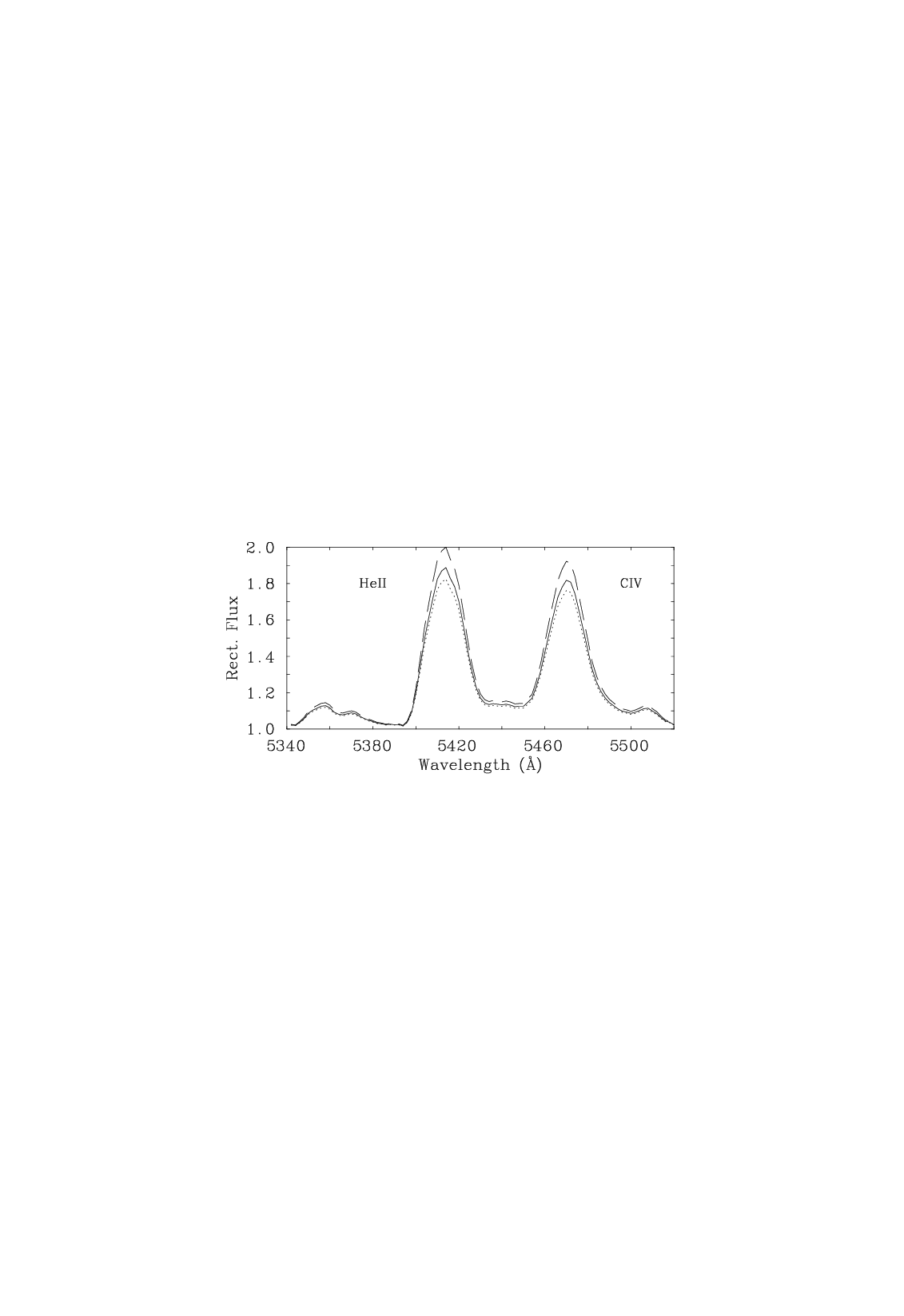}
\caption{Effect of the light ratio uncertainty 
on the strength of important diagnostic lines at He\,{\sc ii} $\lambda$5412
and C\,{\sc iv} $\lambda$5471. (This does not include errors on the 
synthetic atmospheres used in the construction of the wavelength-dependent 
ratio). The dashed and dotted curves correspond to the upper and 
lower error bars, respectively}
\label{light_ratio_errs}
\end{figure}

The spectrum of $\gamma$~Velorum used here is the same phase-averaged rectified
spectrum used in Paper~I. Correction for the continuum as well as the line
contribution of the O star was carried out in that paper using the model O
star spectrum and the deduced value of the light ratio. 
A light ratio of $L$(O) / 
$L$(O+WR) = 0.795 $\pm $0.020 was obtained in Paper~I from 
an average of three
lines, namely \hei\ \la 4471, \hei\ \la 4922 and \heii\ \la 4541. 
Assuming that this light ratio is representative for this spectral range, 
we derive the wavelength dependence of the light ratio using synthetic
spectra for the O star from Paper~I, and for the WR star from the following
section. The ratio of these continuum distributions, adjusted such that
$L_\lambda$(O) / $L_\lambda$(WR) = 3.88$^{+0.5}_{-0.4}$ 
at \la = 4700~\AA,
is plotted in Fig.~\ref{light_ratio}, together with error bars.
A calibration curve $L_\lambda$(O+WR) / $L_\lambda$(WR) is created and 
applied to the optical spectrum of WR11 to account for the contribution
of the O star continuum to the WR equivalent widths. 
In Fig.~\ref{light_ratio_errs} we show the effect that the error
bars from the light ratio have on He\,{\sc ii} $\lambda$5412 and
C\,{\sc iv} $\lambda$5471.


After the spectrum is rectified and light-ratio corrected, 
a model can be calculated to fit the WR lines, as discussed
in the following section. First, we preempt the result of this
modelling in order to verify our choice of reddening.
Published broad-band Johnson V magnitudes for $\gamma$~Velorum span 
a wide range,
from V = 1.70 (Cousins 1972) to V = 1.83 (Johnson et al. 1966). 
We suspect that the broad band observations suffer from
contaminations
by nearby spectral lines such as \civ\ \la 5806 and that
the measured brightness depends crucially on the exact sensitivity
curve of the used setup. 
Therefore, in order to 
minimise the contamination of WR line emission, we prefer to
adopt the narrow-band $ubv$ Smith (1968) photometry. Following the
corrections by Schmutz \& Vacca (1991) to the Smith datasets,
we obtain $v = 1.70$~mag and $b-v = -0.32$~mag. Our theoretical combined O+WR 
model reveals $(b-v)_0 = -0.33$~mag, implying $E(b-v) = 0.01$ or 
$E(B-V)$ = 0.01 mag. 

\begin{figure}
\vspace{9cm}
\includegraphics{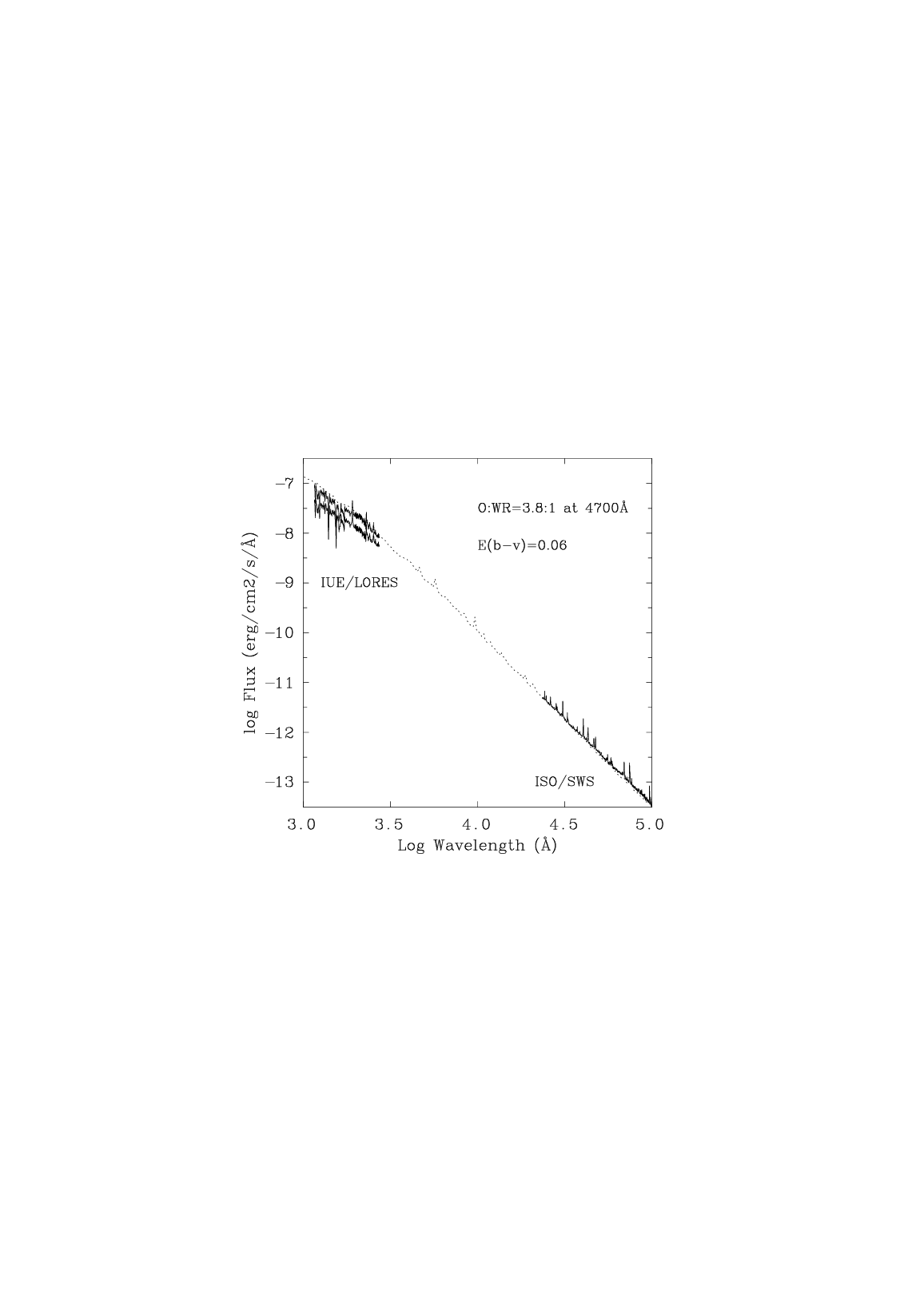}
\caption{Comparison of the combined O+WR synthetic model (dotted lines)
and the IUE/LORES and ISO/SWS observations (solid lines) for reddenings 
of $E(b-v)$ = 0.00 and 0.06 mag, with a light ratio 
$L_{\lambda 4700}$(O)/$L_{\lambda 4700}$(WR) = 3.8}
\label{fig:red}
\end{figure}

We have also carried out an alternative reddening determination, by
fitting the combined WR+O model to UV (IUE/LORES)
and IR (ISO) spectrophotometry, de-reddened following Seaton (1979).
Fig.~\ref{fig:red} compares observed and de-reddened spectrophotometry (solid
lines) with theoretical models (dotted lines), revealing $E(b-v) = 0.06$
or $E(B-V)$ = 0.07 mag. Unfortunately, we are unable to reconcile these 
discrepant reddening values, and so adopt $E(b-v) = 0.035$~mag ($E(B-V)$ = 0.04),
in good agreement with van der Hucht et al. (1988). Note that for this 
reddening and the adopted light ratio, the combined O+WR model flux 
lies $\sim$10\% higher than the UV IUE observed fluxes and a similar 
value below the IR ISO observations.
 
The distance modulus of 7.06~mag (corresponding to a distance of
[$258^{+41}_{-31}$]~pc, 
{\sc hipparcos}\footnote{The {\sc hipparcos}
distance to $\gamma$Velorum has been recently
put in doubt (Pozzo et al. 2000) by the finding of a new population of pre-main
sequence stars which might be associated with the binary. This new
value of the distance re-positions $\gamma$~Velorum within the Vela
OB2 association at a distance of 360-490~pc, in agreement with earlier 
estimates. Clearly, adopting a larger
distance would have a dramatic effect on distance dependent parameters, such as
luminosity, radius and mass-loss (for a distance of 360-490~pc,
$\Delta \log (L/\lsun )$=0.3-0.6). 
We have, however,
maintained the 7.3$\sigma$ {\sc hipparcos} distance measurement, which 
we trust to be the best distance estimate to date.}
and de-reddened 
$v$ = 1.56 mag imply $M_v$(O+WR) = $-$5.50. The O/WR continuum light ratio 
is 3.61 at the $v$-band (Fig.~\ref{light_ratio}) so that $M_v$(O)=--5.23 
and $M_v$(WR) = --3.84. We may convert the Smith (1968) $v$ 
to Johnson $V$ via $V = v - 0.365 (b-v) + 0.007$, adapted from 
Turner (1982) to take account of the $ubv$ re-calibration by 
Schmutz \& Vacca (1991). From theoretical continuum O and WR 
distributions we obtain $(b-v) = -0.34$ and $(b-v) = -0.20$, respectively, 
revealing 
absolute line free V-magnitudes listed in Table~\ref{tab:summary}.
Relative to Paper~I, the difference in the absolute magnitude of the O star, 
$M_V$(O), is $-0.04$ magnitude.
The errors listed in Table~\ref{tab:summary} are solely from the
reddening and light ratio uncertainties. Errors in the distance, 
would add an additional 0.3~mag in the M$_V$ uncertainties.
Propagating these errors onto the derived magnitudes and
luminosities in Table~\ref{tab:summary}, and subsequently onto the 
spectroscopic properties in Sect.~\ref{sec:res}, introduces an additional
systematic error of $\log~(L/L_{\odot})$=$\pm$0.1 for both cases.
In this way we lose sight of the actual accuracy reached by this study.

To be consistent with the slight change in the
O and WR star $V$ magnitudes 
we have revised the uncertainties on $L$, $\cal{M}$
and $i$, while the values of the luminosity and mass of the WR star are
also slightly modified.

\begin{table}
\begin{center}
\caption{O and WR and system parameters,
updated from Paper~I. From these the mass and luminosity of the
WR star are determined. Errors on the distance are not taken into account - 
see text}
\begin{tabular}{lc}
\hline
$\Delta M_{\lambda 4700}$ &  1.47$\pm$ 0.13 mag  \\
$f_{\lambda 4700}$(O)/$f_{\lambda 4700}$(O+WR) &  0.795 $\pm$ 0.020\\
$M_V$(O+WR)     & --5.5  $\pm$ 0.3 mag\\
$i$             &  63$\pm$3 deg            \\
\hline
$M_V$(O)        & --5.10 $\pm$ 0.1 mag\\
$\teff$(O)      &  35\,000$\pm$300 K\\
N[He](O)        &  0.087 (by number)           \\
$R$(O)          &  12.4$\pm$1.7 \rsun    \\
$L$(O)          &  (2.1$\pm$0.3)$\times$10$^5$ \lsun\\
$\cal M$(O)     &  30$\pm$2   \msun        \\
$\mdot$(O)      & (1.78$\pm$0.37)$\times$10$^{-7}$ \msun yr$^{-1}$\\
$\vinf$(O) &  2500$\pm$250 km~s$^{-1}$              \\
age(O)           &  (3.59$\pm$0.16)$\times$10$^6$ yr  \\
\hline
$M_V$(WR)       & --3.76 $\pm$ 0.2 mag\\
$\cal M$(WR) &  9.5$\pm$1.0 \msun      \\
$L$(WR)      &  (1.7$\pm$0.4)$\times$10$^5$ \lsun \\
\hline
\end{tabular}
\label{tab:summary}
\end{center}
\end{table}

\section{Stellar atmosphere codes}\label{sec:mod}

In this section we present a summary of the description of both codes used
(a full account can be found in Hillier, 1989 and Hillier \& Miller, 1998 for
{\sc cmfgen} and in de Koter et al., 1993, 1997, for
{\sc isa-wind}). Along with the description of the basic codes, we
give an account of the Monte Carlo line blanketing code used in 
conjunction with the un-blanketed 
\isa\ to calculate the additional opacity due
to the effect of bound-bound transitions on the continuum photons. Last,
the implementation of the photon loss mechanism (Schmutz 1997)
into the Sobolev code is presented here, for the first time, and a parallel is
drawn with the potential effect of line-line interaction in the
line-blanketed code \cmf.


\subsection{\cmf }
\cmf\
(Hillier 1987, 1990; Hillier \& Miller 1998, 1999) solves the
transfer equation in the co-moving frame, subject to statistical
and radiative equilibrium, assuming an expanding, spherically-symmetric,
homogeneous or clumped, time-independent atmosphere. Line blanketing is
treated correctly in the transfer problem, except that
a simplifying `super level' approach is used (Anderson 1989),
by which several levels of similar energies and properties
are treated as a single `super level'.
The population of an individual atomic level in the full model atom is
determined by assuming that it has the same departure coefficient as
the corresponding super level to which it belongs.
For the specific
models dealt with in this paper the atomic model is shown in 
Table~\ref{table2}. Elemental abundances
other than hydrogen, helium, carbon and oxygen, are fixed at
their solar mass fraction values.

The stellar radius ($R_{\ast}$) is defined
as the inner boundary of the model atmosphere and is located at
Rosseland optical depth of $\sim$20 with the stellar temperature ($T_{\ast}$)
defined by the usual Stefan-Boltzmann relation. Similarly, the effective
temperature ($T_{\rm eff}$) relates to the radius ($R_{\rm 2/3}$)
at which the Rosseland optical depth equals 2/3.
There is now overwhelming evidence for the clumped nature of WR stars
(e.g. Moffat 1999), so we have adopted a simple filling factor
approach. Following the method proposed by Schmutz (1995),
we assume that the wind is clumped with a volume filling factor, $f$, and that
there is no inter clump material. Since radiation instabilities are not expected
to be important in the inner wind we parametrise the filling factor 
so that it approaches unity at small velocities. 
Clumped and non-clumped spectra are very similar, except that line profiles 
are slightly narrower with weaker electron 
scattering wings in the former. Although 
non-clumped models can be easily rejected, we are unable to determine 
the clumping factor because of the severe line blending in WC winds. 
We can therefore derive only $\dot{M}/\sqrt{f}$ with our spectroscopic 
analysis.

Unfortunately, individual line blanketed co-moving frame
calculations are generally computationally demanding, despite
the use of super levels, so that
a large parameter space cannot be quickly explored. One
solution is to (i) solve the transfer problem in the classical Sobolev
approximation rather than the co-moving frame, which the
code {\sc isa}-wind does (de~Koter et al. 1993, 1997),
and (ii) consider
line blanketing via Monte Carlo sampling following 
Schmutz (1991, 1997) allowing the opacity of a huge number of lines to be
considered.  


\subsection{\isa\ and the Monte Carlo line-blanketing code}
The improved Sobolev approximation code ({\sc isa-wind}) is described in
detail by de~Koter et al. (1993, 1997).
The principal differences with {\sc cmfgen} relate to
(a) the treatment of the line radiation transfer; (b) wind
electron temperature which assumes a grey atmosphere in local
thermodynamic equilibrium (LTE); 
(c) the specific atomic model
treated, as listed in Table~\ref{table2}.

The co-moving frame method consistently treats a possible
change of properties of the medium inside the region in which
line photons can be absorbed and re-emitted. In the Sobolev
approximation one assumes this line interaction region is
infinitely small. More exactly, the Sobolev approximation 
assumes that the source function
and the opacities are linear functions such that they can be taken out
of the transfer integral.
This is generally a valid assumption for
Wolf-Rayet stars, where the large velocity gradient of the
flow results in a small interaction zone. The Sobolev
classical approximation (Castor 1970)
implies that interactions of line photons with the continuum
inside of the (in reality extended) interaction region are
ignored. However, the improved Sobolev approximation code
{\sc isa-wind} does include such continuum interactions.
Note that use of the Sobolev approximation introduces great
simplifications in the radiative transfer, resulting in an overall
iteration process up to about 20 times faster compared to
the co-moving frame approach.

Turning to line blanketing, an iterative
technique including the Monte Carlo method of
Schmutz (1991) is employed.
The method allows the computation of intensity-weighed
effective opacity factors, which account for the presence of tens of
thousands of spectral lines, dominated by Fe and Ni. Based on
{\sc isa-wind} atmosphere
calculations, the Monte Carlo code determines the line blanketing
factors. An iterative procedure is used, such that blanketing factors
are used by the non-LTE code to calculate a new atmosphere,  which
in turn is used to calculate new blanketing factors. A few iterations
are generally sufficient. This is due to the fact that the scattering
and absorption factors are not very sensitive to the exact model intensity.

The Monte Carlo method deals with the photon scattering in the correct manner,
although no branching is taken into consideration, an effect that would tend to re-distribute
UV photons to longer wavelengths.
The ionization equilibrium of metal species is also approximate,
in that it is determined in relation to the non-LTE structure of
helium, carbon and oxygen and not by the full solution of the statistical
equilibrium equations for that species.


\subsection{Photon loss}\label{ssec:pholos}

``Photon loss" is the name given by Schmutz (1997) to the
interaction between the photon field of the He~{\sc ii} 
Ly$\alpha$ line at 303.78~\AA\
and nearby metal lines (e.g. the \oiii\ Bowen lines at
305.72 and 303.65~\AA ).
This was suggested by Schmutz (1997) to be responsible for a decrease
in the ionization equilibrium of the wind of WR6 (WN4), with respect 
to the equilibrium obtained when no such 
line-line interaction was accounted for. 
In order to fit the observed spectrum when photon loss is included in the
model, the model's stellar temperature 
has to be higher, which results in a higher  
spectroscopic luminosity. This contributes to
a lower wind performance number and allowed Schmutz (1997) to calculate a
velocity structure for the outer wind of WR6.

Since line-line interaction was not accounted for in any
atmospheric wind code before Hillier \& Miller (1998) implemented it into the
Hillier (1990) code, Schmutz (1997) calculated the photon loss
factor (i.e. the amount of interaction between the \heii\ Ly$\alpha$
and nearby metal lines) by comparing the opacity of metal lines
in the 303-\AA\ region
to that of the \heii\ Ly$\alpha$ line 
(his Appendix A). Once the amount of 
interaction is known for a particular model atmosphere, a corresponding
fraction of photons can
be taken out of the radiation field at that particular wavelength
for all grid points. In this way the same fraction of photons
is removed from the radiation field at all depth points, although in
reality the
photon loss factor depends on the relative opacity of the \heii\ 
Ly$\alpha$ line and
that of the metal lines, which in turn depends on the depth-dependent
populations. This problem however is not critical
as long as the opacity of metal lines behaves similarly to that of helium.
Additionally, the loss of photons is only critical for the point
in the atmosphere where the wind is recombining. 
We implemented the photon loss approximation
into the \isa\ code of de~Koter et al. (1993 - see Sect.~4.3.1)
and used it in the calculation of a model for WR11. 

\cmf\ includes line-line interaction and as such it should naturally 
include the photon 
loss effect, provided the specific metal lines in the \heii\ Ly$\alpha$ region 
are included in the model atom. We can therefore
test whether this effect and its consequences for the stellar luminosity
has been accounted for properly by our approximation. 


\subsubsection{Implementation of Photon loss in \isa }

Once the photon loss factor, $f$, is calculated for a certain model atmosphere
(see Schmutz 1997, Appendix A),
a fraction $f$ of the photons is removed from the radiation field
in the wavelength range 300--304\AA . To do so, one has to modify
the expression where the mean intensity for this line wavelength
is calculated at every code iteration.
In \isa\ this was achieved via a modification of the Sobolev escape 
probability.

In the Sobolev approximation the radiation field $\bar J$ at
radius point $r$ is related to the line source function
$S$ in the following way:

\begin{equation}
\bar J(r) = (1 - \beta ) S(r) 
\end{equation}

\noindent where $\beta$ is the line escape probability by direct flight 
and where

\begin{equation}
S(r) = {N_u A_{ul} \over N_lB_{ul} - N_u B_{ul}}
\end{equation}

\noindent with $N_u$ and $N_l$ the upper and lower level populations,
respectively, $A_{ul}$ and B$_{ul}$ the Einstein coefficients.
Due to photon loss, the field at frequencies close to the
He~{\sc ii} Ly$\alpha$ is modified in the following way:

\begin{equation}
\bar J(r) \rightarrow (1 - f) \bar J (r) .
\end{equation}

\noindent Substituting (1) into (3) we obtain:

 
\begin{equation}
\bar J (r) = (1 - \beta -f +f\beta) S(r) 
\end{equation}

\noindent Comparing (5) to (1) we see that the classical Sobolev escape 
probability is modified in the following way:

\[
\beta \rightarrow (\beta + f -f\beta ). 
\]


\section{Stellar analysis}\label{sec:res}

In this section we present the stellar parameters for WR11
derived with the two codes \cmf\ and \isa\ (supplemented by
the Monte Carlo line-blanketing code). We will discuss the quality of the fits
and the overall flux distribution of the model atmospheres.


\subsection{Analysis technique}

In our approach, diagnostic
optical lines of He\,{\sc i} ($\lambda$$\lambda$5876, 10830) and
He\,{\sc ii} ($\lambda$$\lambda$4686,5412) 
are chosen to derive
the stellar temperature and mass-loss rate.
Initially we assume the carbon and oxygen abundances to be the same 
as those derived for the single WC8 star
WR135 by Dessart et al. (2000), because of the similarity of the
two spectra (a comparison of the rectified spectra is shown in 
Fig.\ref{wr135_fig}). 
Once a fair agreement is reached between
the modelled and the observed helium lines, 
carbon and oxygen abundances can be altered to improve
the fits to their diagnostic lines. For carbon, lines of
\ciii\ at 6741~\AA\ and of \civ\ at 5471~\AA\ are used.
For oxygen, the lack of good diagnostic lines in our
spectral range and the weak and blended nature of those that can be identified,
make the determination of the oxygen abundance impossible. We therefore
adopt the theoretical O/C number ratio of 0.2 (Meynet et al. 1994).
Changing oxygen and carbon abundances acts on the total opacity of the wind and therefore
feeds back on the modelled strengths of the helium lines. 
Parameters have therefore to be adjusted 
so as to retrieve the fits of the helium lines.

We adopt the wind terminal velocity determined by St.Louis et al. 
(1993), $v_\infty = 1550$~km~s$^{-1}$,
from the variability of the UV spectrum. 
The velocity law used by the \cmf\ code (Hillier \& Miller 1999)
is a two component ``beta" law which produces a slower flow
than the usual $\beta$ = 1 velocity law,
at points in the atmosphere intermediate between the outer regions
where $v \rightarrow v_{\infty}$ and the photosphere. 
Hillier \& Miller (1999)
found that it leads to better fits to the line profiles, while at the same
time Schmutz (1997) determined a similar velocity law from his
hydrodynamic calculation (which, however was carried out only for the
outer layers of the wind). The two component velocity law 
used for our calculations
uses $v_0$ = 100~km~s$^{-1}$ (the photospheric velocity), 
$v_{\rm core}$ = 1.0~km~s$^{-1}$ (the velocity at $R_*$), 
$v_{\rm ext}$ = 1100~km~s$^{-1}$ (the intermediate velocity) and 
$v_\infty$ = 1550~km~s$^{-1}$ (the terminal velocity). The two beta
exponents used are $1.0$ and $50$, for $\beta_1$ and $\beta_2$, respectively.
The \isa\ calculation was performed with the same velocity law as \cmf .
A clumping filling factor of 0.1 was adopted for \cmf . Usually this 
is derived together with the other parameters,
by fitting line wings; for WC stars the heavy blending makes this
task harder. We therefore adopted the value obtained by
Hillier \& Miller (1999) and confirmed by Dessart et al. (2000). 

The atomic data models used for the two codes are summarised in 
Table~\ref{table2} (see Dessart et al. (2000) for a description of the
atomic data used in \cmf , while \isa\ uses the latest Opacity Project data). 
Note that the number of 
individual and super levels ($N_F$ and $N_S$) mean different things
for the two codes: 
\cmf\ bundles $N_F$ levels into $N_S$ super-levels, while all \isa\ atomic levels,
listed under the $N_F$ heading,
are treated individually (apart for \hei\ and \civ\ for which the upper
32 and 6 levels, respectively, are super-levels).
The data used by {\sc cmfgen} is more extensive than 
those used by {\sc isa-wind}. Generally this can have an effect on the
ionization equilibrium (and hence on the spectrum), or only
have an effect on particular lines between high lying levels. Tests using
\cmf\ with reduced atomic data show that while individual lines can be very sensitive
to the amount of levels used (e.g. \ciii\ $\lambda 8500$) the overall
ionization of the wind is not critically dependent on the size of the atomic model
provided a minimum amount of levels is used.

\begin{table*}
\caption{Summary of model atom used in {\sc cmfgen} and {\sc isa}-wind
radiative transfer calculations, including full levels ($N_{\rm F}$),
super levels ($N_{\rm S}$)
and total number of transitions ($N_{\rm Trans}$), following
Hillier \& Miller (1998) and de~Koter et al. (1997; 31 auto-ionizing levels are
included for \ciii\ in the atomic data used by \isa )}
\label{table2}
\begin{center}
\begin{tabular}{lrrrrrll}
\hline\noalign{\smallskip}
Ion & \multicolumn{1}{c}{$N_{\rm S}$} & \multicolumn{2}{c}{$N_{\rm F}$}
& \multicolumn{2}{c}{$N_{\rm Trans}$} & \multicolumn{2}{c}{Details} \\
   & {\sc cmfgen} & {\sc cmfgen} & {\sc isa}-wind &
     {\sc cmfgen} & {\sc isa}-wind & {\sc cmfgen} & {\sc isa}-wind \\
\hline\noalign{\smallskip}
He\,{\sc i}   & 27 & 39 & 51 & 905 &588 &$n \le$14          & $n\le$20 \\ 
He\,{\sc ii}  & 13 & 30 & 20 & 435 &190 &$n \le$30.         & $n\le$20 \\ 
C\,{\sc ii }  & 39 & 88 & 7  & 791 & 10 &$nl \le$2s2p3p$^2$D& $nl \le$2s$^2$($^1$S)3d$^2$D \\
C\,{\sc iii}  & 99 &243 & 44 &5513 &174 &$nl \le$2s10z$^1$Z & $nl \le$2s($^2$S)5f$^1$F$^{\circ}$\\
C\,{\sc iv}   & 49 & 64 & 15 &1446 &105 &$n \le$30          & $n\le$10 \\ 
O\,{\sc ii}   & 23 & 75 & -- &   3 &--  &$nl \le$2p$^3$$^2$P$^o$                  & --      \\
O\,{\sc iii}  & 50 & 50 & 3  &  213 &--  &$nl \le$2p3d$^1$D$^o$                    & 2s$^2$2p$^2$($^1$S)          \\
O\,{\sc iv}   & 30 & 72 & 22 &  67& 52 &$nl \le$2s2p4d$^2$P$^o$                  & 2s2p($^3$P$^{\circ}$)3d$^4$D \\
O\,{\sc v }   & 31 & 91 & 24 & 835 & 57 &$nl \le$2s4f$^1$F$^o$                    & 2p($^2$P)3p$^1$S        \\
O\,{\sc vi}   & 13 & 19 & 26 & 749 &153 &$nl \le$8d$^2$D                          & 1s$^2$($^1$S)10d$^2$D        \\
Si\,{\sc iv}  & 12 & 20 & -- &  72 & -- &$n \le 20$                               &-- \\
Ca\,{\sc iii} & 16 & 39 & -- & 104 & -- &$nl \le$3s$^2$3p$^5$5s$^3$P$^o$          &-- \\
Ca\,{\sc iv}  &  8 & 26 & -- &  12 & -- &$nl \le$3s$^2$3p$^4$($^1$D)3d$^2$D       &-- \\
Ca\,{\sc v}   & 19 & 32 & -- &  44 & -- &$nl \le$3s3p$^4$($^4$P)3d$^5$D           &-- \\
Ca\,{\sc vi}  & 16 & 30 & -- &  61 & -- &$nl \le$3s3p$^3$($^3$D)3d$^4$G$^o$       &-- \\
Fe\,{\sc iv}  & 21 &280 & -- &3767 & -- &$nl \le$3d$^4$($^1$G$^o$)4p$^2$P$^o$      &-- \\
Fe\,{\sc v}   & 19 &182 & -- &1921 & -- &$nl \le$3d$^3$($^2$D$^o$)4p$^1$P$^o$     &-- \\
Fe\,{\sc vi}  & 10 &80  & -- & 617 & -- &$nl \le$3d$^2$($^1$S)4p$^2$P$^o$         &-- \\
\noalign{\smallskip}\hline
\end{tabular}
\end{center}
\end{table*}


\subsection{Optical and Infra-red line fits}\label{ssec:fits}

\begin{figure*}
\vspace{20cm}
\includegraphics{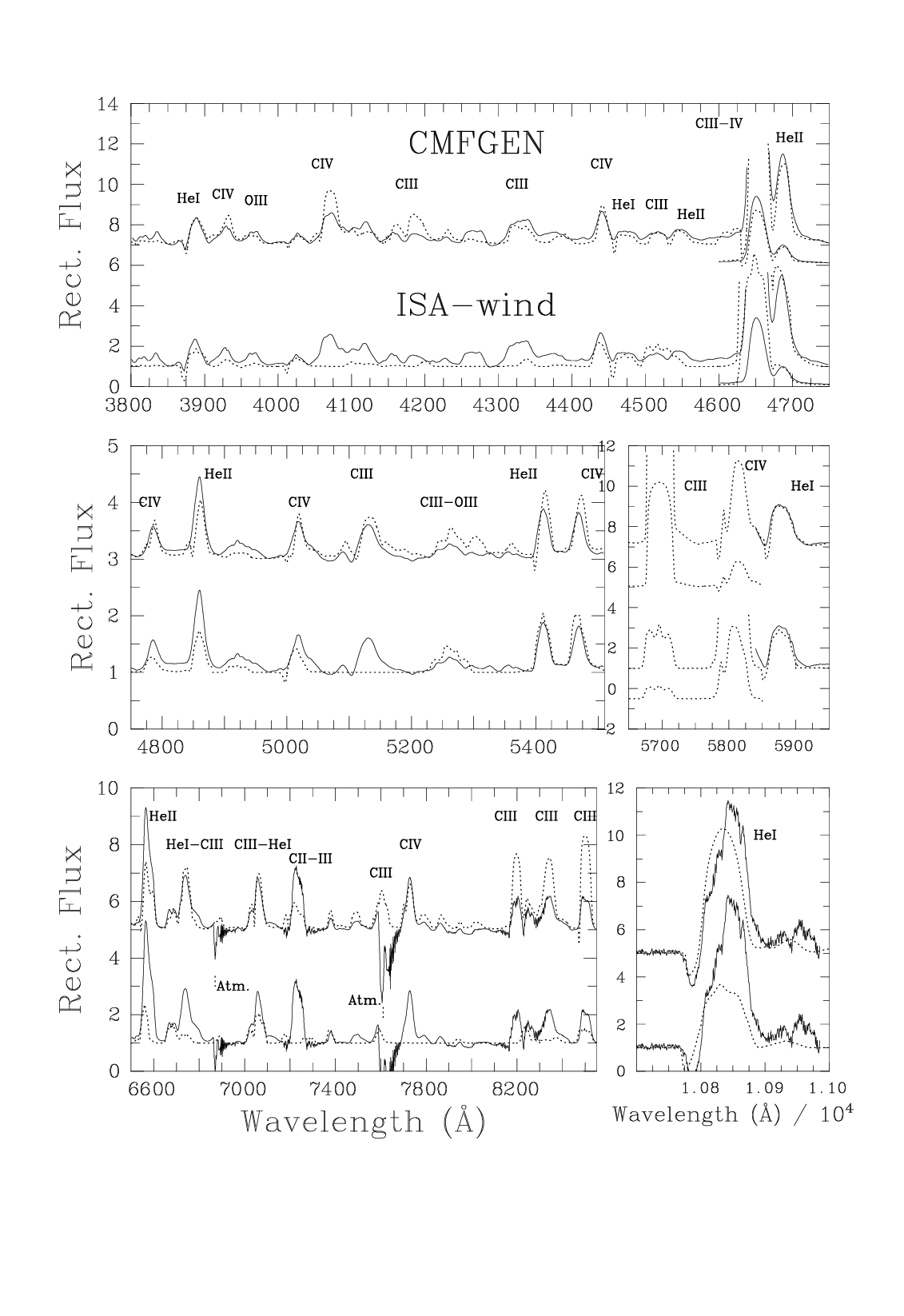}
\caption{Comparison between the rectified optical ({\sc heros}) 
spectrum of WR11 (solid line) with the synthetic spectrum
produced by \cmf\ (upper spectrum, dotted line) and
{\sc isa}-wind (accounting for line blanketing - lower spectrum, 
dotted line). For clarity of display, the regions 4600-4700~\AA\
and 5650-5850~\AA\ are multiplied by 0.2 and 0.3, respectively, and offset}
\label{fig2}
\end{figure*}

The overall fit quality of the helium diagnostic lines is fair and 
comparable for the two codes (Fig.~\ref{fig2}). Due to the difficulties
in rectifying WC spectra and the weakness of the lines in the region
4400--5450~\AA\ region, it was decided that the
\hei~\la \la 5876,10830 would be better diagnostic lines than \hei\
\la 4471, although the fit to this line is not inconsistent. \la 5876 is very well
reproduced by \cmf , while \la 10830 is under-predicted by about 15\% 
(one should take into account that
this line was not corrected for the O star absorption line,
which can be observed as a depression on its blue side). 
\isa\ only reproduces \la 5876 (albeit
not perfectly), due to the
sensitivity of \la 10830 to the outer wind electron temperature, which
is calculated in the grey LTE approximation and fixed for outer radii to
10\,000~K.

\heii~\la 5412 is well reproduced by both codes, although \cmf\
slightly over-predicts it. Once again we use the 4440-4600~\AA\
region only as a secondary diagnostic: \cmf\ does reproduce the \heii\ \la
4541-\civ\ \la 4553 rather well, while \isa\ does not predict
the \civ\ component in the blend with an overall poorer result.
\la 4686 is slightly under-predicted by \cmf , with the \isa\ 
synthetic line fit being affected by blending with the over-predicted
\ciii -\civ\ line at 4660~\AA .

\begin{figure*}
\vspace{15cm}
\includegraphics{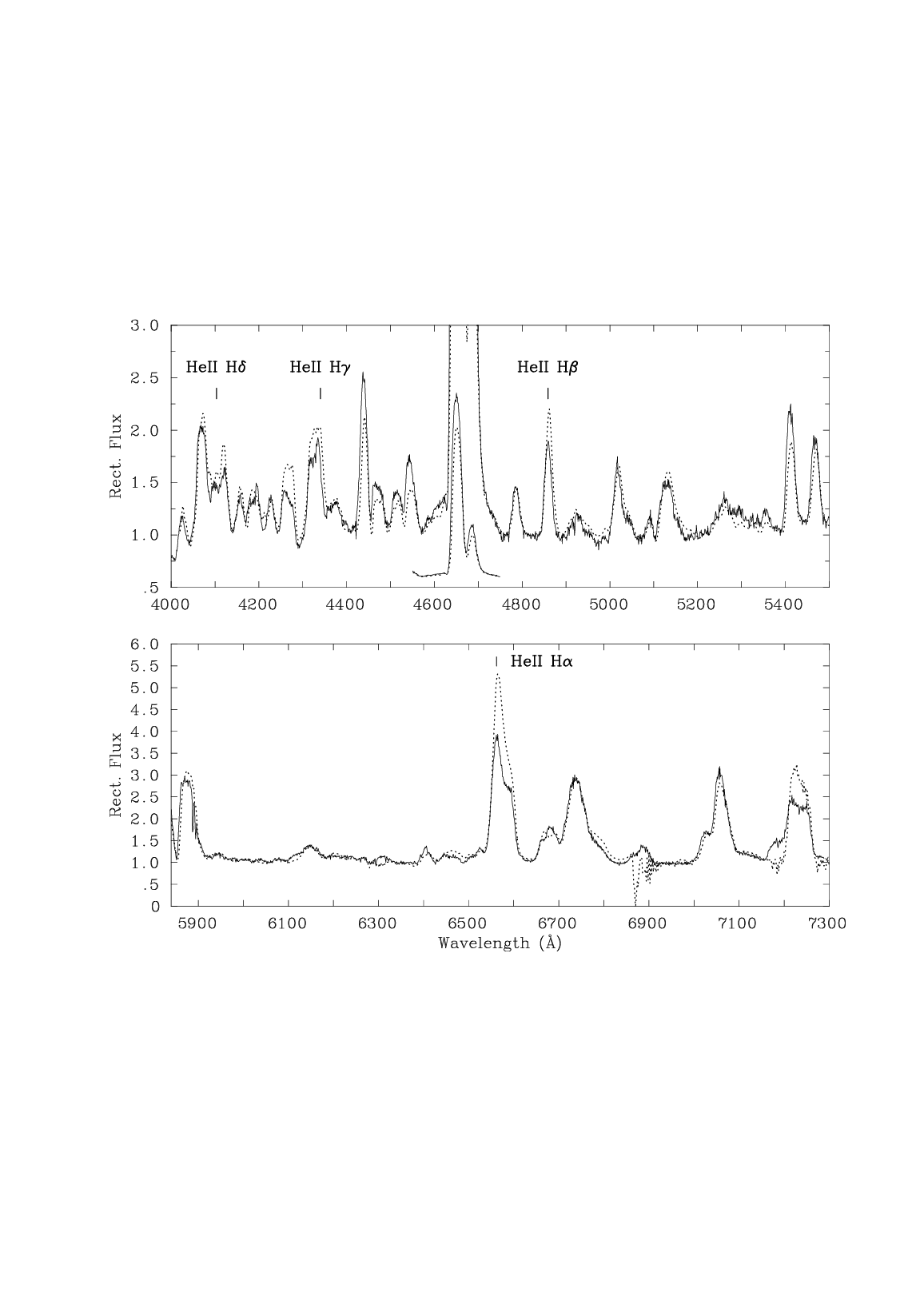}
\caption{Comparison between WR135 (solid line) and WR11 (dotted line), showing the
similarity of the two spectra except for the \heii\ lines at 4859 and 6560~\AA }
\label{wr135_fig}
\end{figure*}

The \heii\ lines at \la 4859 and \la 6560 \AA\ are underestimated by 
both \cmf and \isa , by about 20\% 
and 80\% 
respectively. In the case of \la 6560
the presence of an under-predicted flat-topped \cii\ line could be
responsible (see below). It is in fact not uncommon that models under-predict
line emission from low ionization stages (cf. \cii $\lambda$4267), showing 
that the stratification
of the wind is imperfectly reproduced. It is possible that this problem will
be solved once more realistic clumping is included in the models.
On the other hand there is no blend in \la 4859 that could be blamed
for the poor fit. For that line there could be a slight rectification problem,
although this alone could not account for the amount by
which the model under-fits the data.

In Fig.~\ref{wr135_fig} we show a comparison of the observed spectra of WR11 and
WR135. 
The two spectra are very similar with WR135 having
only slightly stronger \heii\ \la \la 4541, 4686, 5412 lines 
and comparable \hei\ \la \la 5876,7066 lines.
On the other hand, 
WR11 has stronger \heii\ \la \la 4339, 4859, 6560, 
indicating an opposite trend 
to the rest of the \heii\ spectrum! (\heii\ \la 4100 is also shown in
Fig.\ref{wr135_fig} although due to the heavy blending with carbon lines,
it is difficult to make a clear comparison). This points to an anomaly in the strength
of these \heii\ Pickering lines. We should also note that the ratio of
\heii 5411 to H$\beta$ is different for the two stars, indicating something anomalous 
in the line formation in one or both stars.

For carbon lines,
\cmf\ reproduces strengths and shapes to a higher degree of
accuracy. \isa\ synthetic lines underestimate 
\civ\ \la \la 4440,4780 by about 20\% and 40\%, respectively, while
\ciii\ \la \la 6741,8500 are weaker than the observations by 80\%
and 50\%,
respectively. On the other hand while \cmf\ reproduces  \la \la 4440,4780,6741,
the \ciii\ line at 8500~\AA\ is over-estimated by more than 100\%.
The \la 4660 \ciii\ - \civ\ blend is
overestimated by \isa\ ($\sim$60\%), while it is underestimated by \cmf\
($\sim$10\%).
Some lines of \ciii\ (e.g. \la \la 8196,8500) and \civ\ (\la 7063), 
are under-predicted or not predicted at all by \isa\
partly because they are not accounted
        for in the atomic model. The carbon abundance determination,
however, rests critically on
the ratio of \heii\ \la 5412 / \civ\ \la 5471 which is well matched by both
codes. \cii\ is under-predicted by \cmf\ and not predicted at all
by \isa\ (which includes only 7 levels in the atomic model of \cii ). However
the presence of the \cii\ line at 4267~\AA\ 
indicates that \cii\ lines are present in the spectrum.
The spectral classification lines \ciii\ \la 5696 and \civ\ \la 
5806 are not included
in the observations, so in Fig.~\ref{fig2} we only show the prediction for
both models. These two lines are remarkably different in the two
models: the spectral types deduced from their equivalent width ratio 
are WC6 and WC9 for
the \isa\ and \cmf\ codes, respectively, where the criteria of
Crowther et al. (1998) have been adopted. 
The model for WR135 of Dessart et al. (2000), with similar 
parameters, has a much larger \civ\ \la 5806/\ciii\ \la 5696 line ratio,
consistent with a WC8 classification. 
Although these two lines form an excellent 
spectral type diagnostic because of their mutual proximity and their strengths,
it is important to note that the correspondence of their strengths to the
stellar parameters is not straight forward.

Differences in the carbon spectrum are obtained (De Marco et al. 1999b)
not only between 
\cmf\ and \isa , but also with the co-moving frame code of Koesterke
\& Hamann (1995). Such discrepancies are not due to
differences in the atomic data, and it is extremely difficult to
identify the exact cause, since the effect is highly
dependent on the parameter space investigated. 
Further investigations are under-way.

\begin{figure*}
\vspace{8cm}
\includegraphics{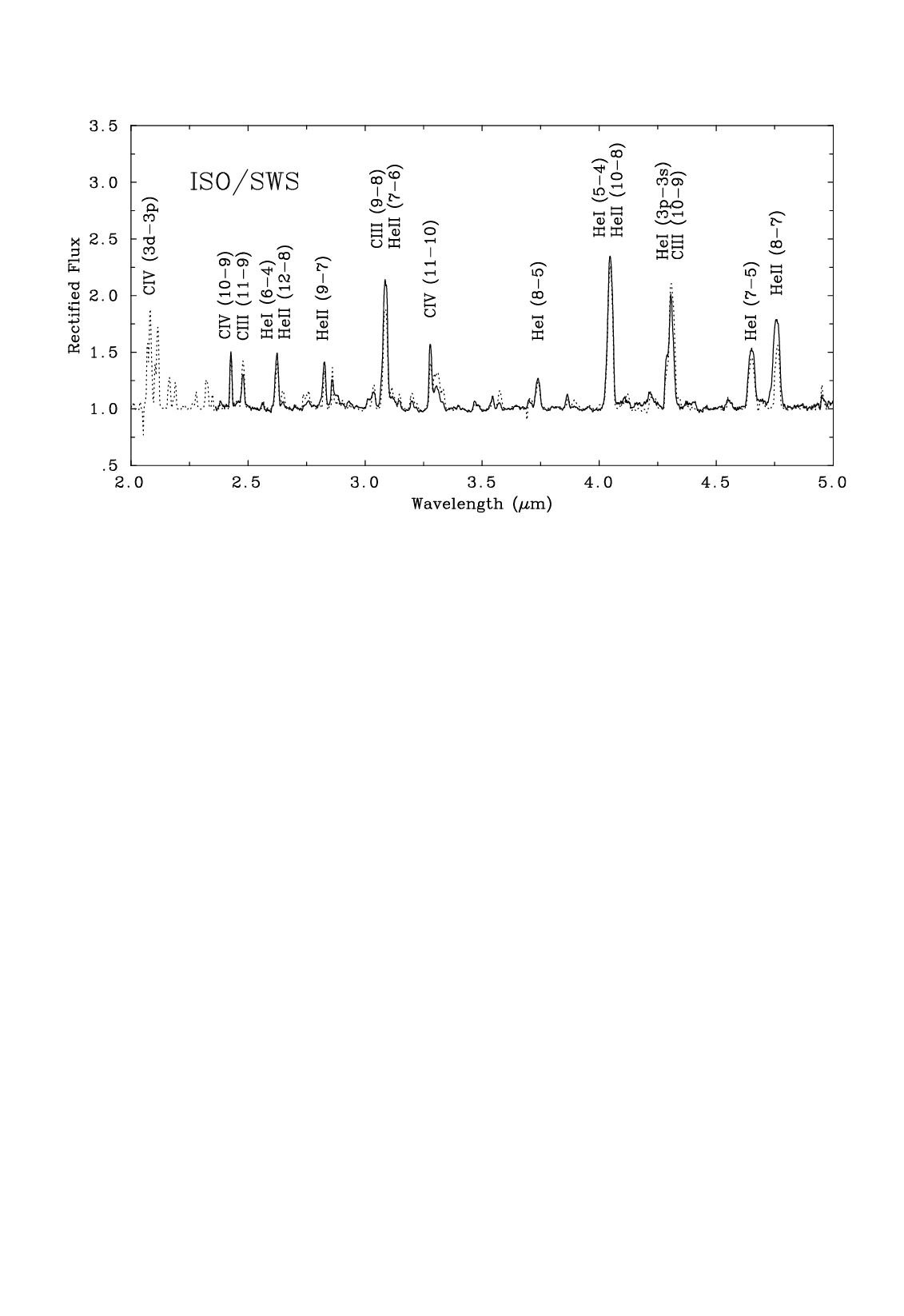}
\caption{Comparison of the rectified ISO observations of $\gamma$~Velorum (solid)
with the \cmf\ model (dotted) scaled to the optical L$_{\lambda 4700}$(O) / 
L$_{\lambda 4700}$(WR) = 3.8}
\label{wr11_wr135_ISO}
\end{figure*}

A comparison of the \cmf\ synthetic spectrum with 
the rectified ISO spectrum of $\gamma$~Velorum
is shown in Fig.~\ref{wr11_wr135_ISO}. Our WR+O model spectrum fits
the emission lines to a good level of accuracy, demonstrating that line
fits are consistent also in the IR.  Spectral features in the H and K
windows agree qualitatively with unpublished AAT spectroscopy of
L.F. Smith (priv. com.). As mentioned in 
Sect.\ref{sec:red}, the combined O+WR model continuum is $\sim$10\%
lower than the flux-calibrated ISO spectrum for the adopted light ratio, 
reddening and observed $v$ magnitude.

In conclusion it was felt that given the limitations of the observed
spectra, the parameters of WR11 from the optical fits have been
constrained rather well. The uncertainties listed in Table~\ref{wr11_results}
are determined from the fits (for the $T_*$, $\dot{M}$, C/O 
and $v_\infty$ parameters) or
from error propagation. In the case of the spectroscopic luminosity, $L$,
the uncertainty derives directly from the uncertainty on $M_V$
(see Sect.~\ref{sec:red}) since the uncertainty on the bolometric correction BC
is negligible. For the radius, $R_*$, 
the error propagates from the luminosity and
temperature uncertainties. Luminosity, mass-loss and terminal velocity
determine the uncertainty on the wind efficiency $\eta$. Finally the
uncertainty on the ionizing fluxes, $Q_0$ and $Q_1$ derive directly
from $L$, without taking into account internal model uncertainties.

Given the different nature of the two non-LTE codes, it is remarkable
how similar the deduced parameters are. This confirms that despite the
approximations adopted by \isa , this faster code is suitable
for the analysis of Wolf-Rayet stars in the WC8 parameter domain 
(as it was also shown to be appropriate for a WN8 star by Crowther et al. 1999).

\begin{table}
\begin{center}
\caption{Comparison of stellar parameters obtained for WR11
using {\sc cmfgen} and {\sc isa}-wind (blanketed including photon loss)
}
\label{wr11_results}
\begin{tabular}{cccc}
\hline
Parameters & {\sc cmfgen} & {\sc isa}-wind & Error \\
Filling factor, $f$ &  10\% &  100\%    & \\
Photon Loss &  -- &  yes & \\
\hline                                   
$T_{\ast}$ [kK]                & 57.1     & 56.0  & 3\% \\
log($\mdot/\sqrt{f}$ [$M_{\odot}$yr$^{-1}$])& --4.53 &--4.48 & 20\%  \\
C/He[by number]                & 0.15     & 0.06  & 40\% \\
O/He[by number]$^a$                & 0.03     & 0.01   & -- \\
R [R$_{\odot}$]                & 3.2      & 3.2    & 20\%\\
log($L_{\ast}/L_{\odot}$)      & 5.00     & 4.95   & 30\% \\
BC [mag]                       & --4.00   & --3.89  & negl.\\
$v_\infty$ [km~s$^{-1}$]       & 1550     & 1550   & 10\% \\
$\eta$/$\sqrt{f}$              & 21       & 28     & 60\% \\
log(Q$_0$/s$^{-1}$)            & 48.81   & 48.75 & 30\%\\
log(Q$_1$/s$^{-1}$)            & 47.76    & 47.83 & 30\% \\
\hline
\multicolumn{4}{l}{$^a$This value follows from the adopted O/C=0.2}
\end{tabular}
\end{center}
\end{table}

\begin{figure}
\vspace{13.5cm}
\includegraphics{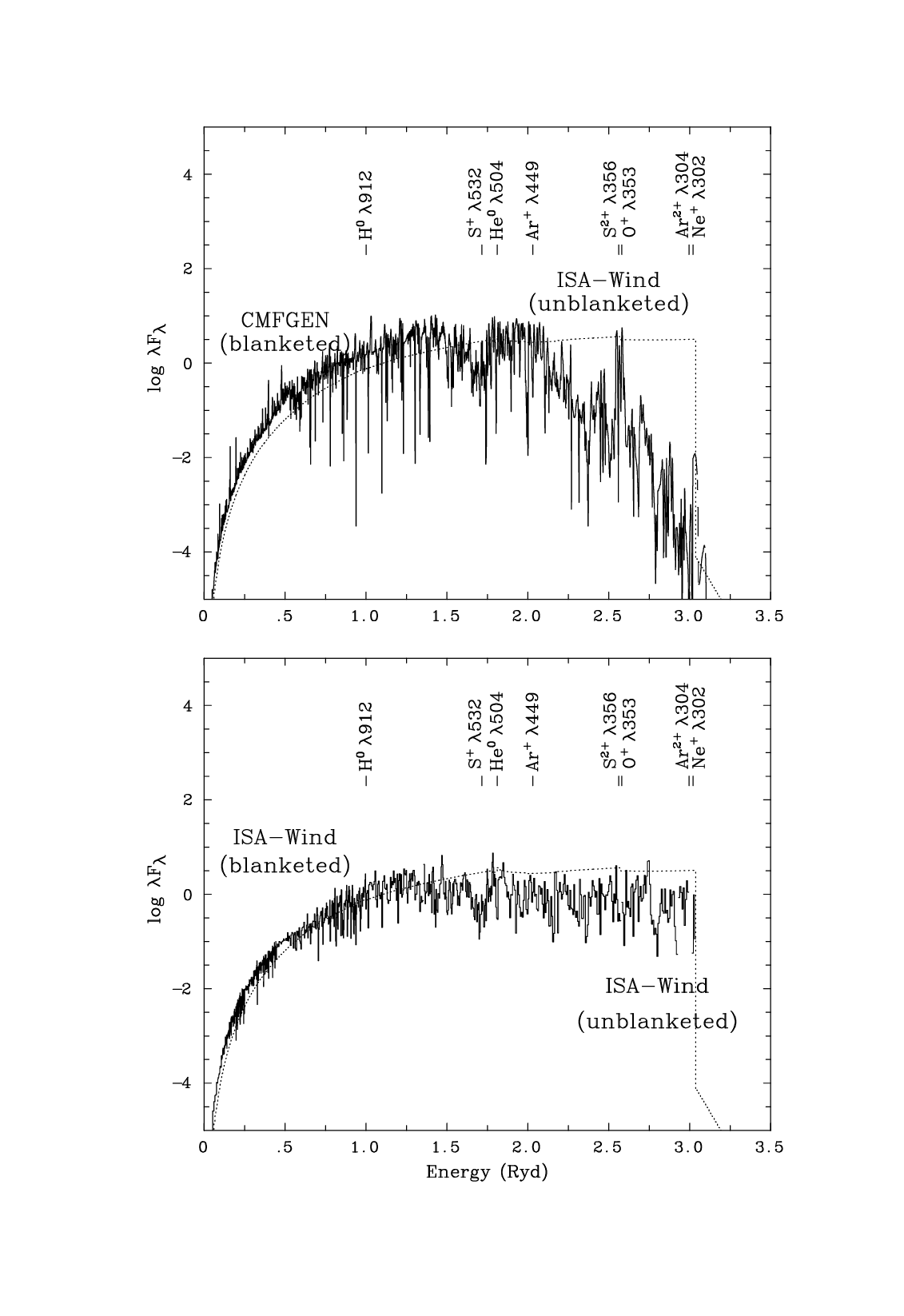}
\caption{Comparison between the output flux of \cmf\ (top) and
{\sc isa}-wind (bottom). Both fluxes are compared to the un-blanketed
\isa\ spectrum for the same stellar parameters}
\label{wr11_flux}
\end{figure}


\subsection{UV ionizing fluxes and bolometric luminosity}

In Fig.~\ref{wr11_flux} we present the predicted extreme UV fluxes from
\cmf\ (solid line, top panel) and \isa\ (solid line, bottom panel) compared
with the un-blanketed \isa\ model flux (dotted lines) on the energy vs
log ($\lambda$F$_\lambda$) plane. 

As can be seen, the \isa\ flux at $\lambda \le 400$~\AA\ 
is harder than the \cmf\ flux, a discrepancy
that was already pointed out by Crowther et al. (1999).
On the other hand the
Rydberg vs. $\lambda F_\lambda$ plot is designed to emphasise the
0-912~\AA\ region (1~Rydberg = 13.6~eV) and we see by looking at
the values of log(Q$_0$/s$^{-1}$) and
log(Q$_1$/s$^{-1}$) shown in Table~\ref{wr11_results}, that
in fact the two fluxes are not particularly 
dissimilar short of 912~\AA\ and 504~\AA . 
On the
other hand, the fluxes shortward of $\sim$400~\AA , which influence 
the strengths of nebular lines such as O$^{2+}$, Ar$^{3+}$ and
Ne$^{2+}$, are very different and this 
discrepancy could affect derived nebular properties.

The spectroscopic
luminosity determined by \cmf\ and \isa\ 
are not dissimilar, although the former is slightly higher 
(Table~\ref{wr11_results}).
Both the luminosities are lower than that determined in Paper~I from
alternative methods (log(L/L$_\odot$) = 5.2$\pm$0.1), even considering the
formal uncertainty determination.

The luminosity of 63000~L$_\odot$ 
obtained by Schaerer et al. (1997) can be compared
to that obtained here by the \cmf\ code of 100\,000~L$_\odot$. We can 
accredit the increase to the presence of line blanketing (which includes photon
loss from the \heii\ Ly$\alpha$ - see next section), although the difference of
the two codes and the neglect of carbon and oxygen might also play a role.

\section{The effect of line-line interaction}\label{sec:pholos}

Line-line interaction
between the very strong \heii\ Ly$\alpha$ and nearby weaker lines
has a large effect on the wind ionization structure. Approximations in the
line-blanketing technique that affect either the position or the
strength of even the weakest lines can affect greatly the derived
stellar parameters. In this section we determine the size of this
effect and show its consequences for the derived parameters of 
$\gamma$~Velorum.

Schmutz (1997) proposed that capturing
of \heii\ Ly$\alpha$ line-photons by resonant metal transitions, would  
reduce the ionization balance of the wind. Including line-line interaction between
\heii\ Ly$\alpha$ and lines from ions such as \cav , \fevi , \niv\ and \oiii\ 
in his non-LTE model of the WN4 star WR6, 
led to a significantly lower wind ionization structure than that obtained
without taking into account such interaction. As a result, 
he had to adopt a higher effective temperature
to fit the spectrum of this star, leading to a larger luminosity and a lower
wind performance number. In this way enough radiation force was obtained 
to drive the outer parts of the wind of WR6.

Schmutz (1997), however, carried out only an approximate calculation, which
kept the fraction
of photons captured by metal lines constant throughout the wind, instead of
varying with radius. Additionally, the populations
of the levels responsible for the metal transitions intercepting the \heii\ photons,
were calculated in LTE, instead of being calculated consistently with the populations
of all other levels via the statistical equilibrium equations.
Since the extent of this effect could
lead to the long-sought answer to a driving mechanism for WR stars, we thought it
important to investigate it further.

The line-blanketed \cmf\ code (Hillier \& Miller 1999) 
automatically takes into account line-line interaction between
all the lines included in the adopted atomic model so that
photon loss should naturally take place 
in our calculation of Sect.~\ref{sec:res},
between \heii\ Ly$\alpha$ and the included lines of
iron, oxygen, calcium and carbon.
To determine the effect of line-line interaction on the wind ionization,
we removed from the atomic model
sets of lines in the spectral region neighboring
($\pm 2000$~km~s$^{-1}$) \heii\ Ly$\alpha$. 
{\it Throughout the rest of this discussion, we will refer to the
model calculated with this reduced atomic model the ``no-PL model"
(which stands for no photon loss), while the model calculated in
Sect.~\ref{sec:res}, which fits the spectrum of $\gamma$~Velorum,
is referred to as the ``basic model"}.

\begin{figure*}
\vspace{11cm}
\includegraphics{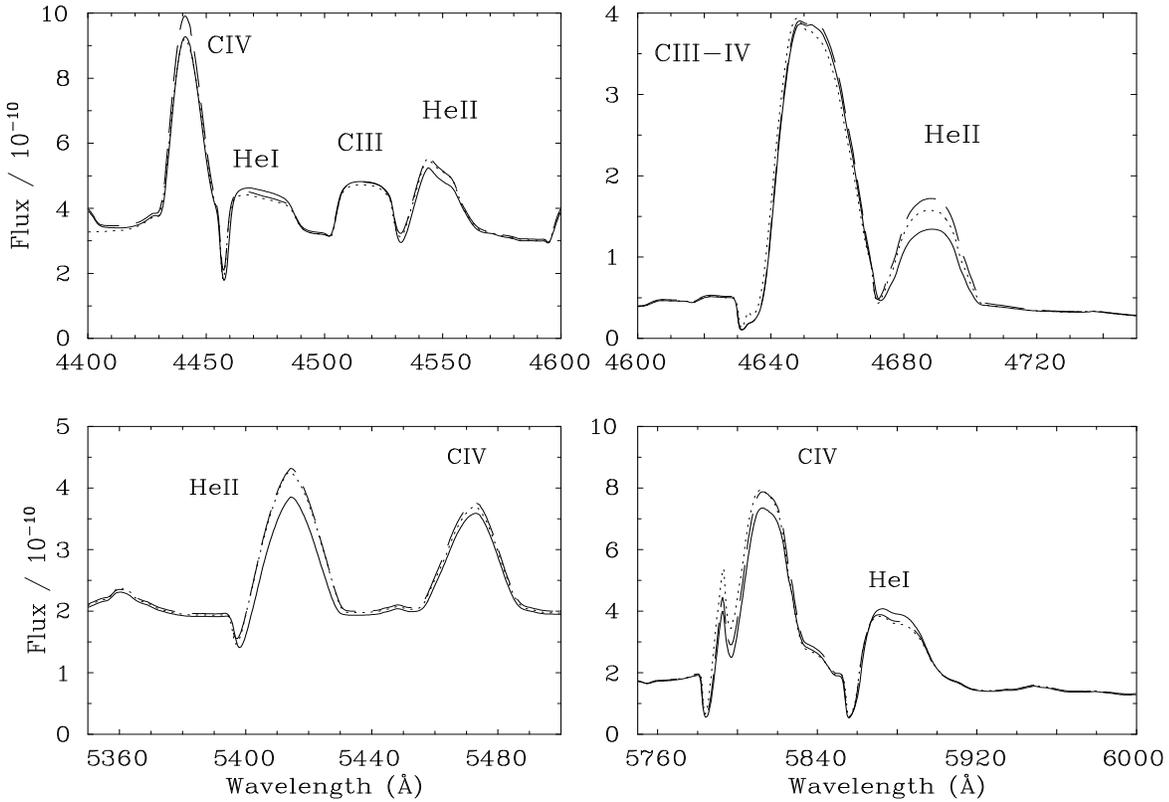}
\caption{Comparison between the model presented in Sect.~\ref{sec:res} (solid), 
the model with metal lines in the \heii\ Ly$\alpha$ region removed (dashed)
and the model with the reduced $v_{\rm turb}$ (dotted)}
\label{all_fluxes}
\end{figure*}

\begin{figure*}
\vspace{11cm}
\includegraphics{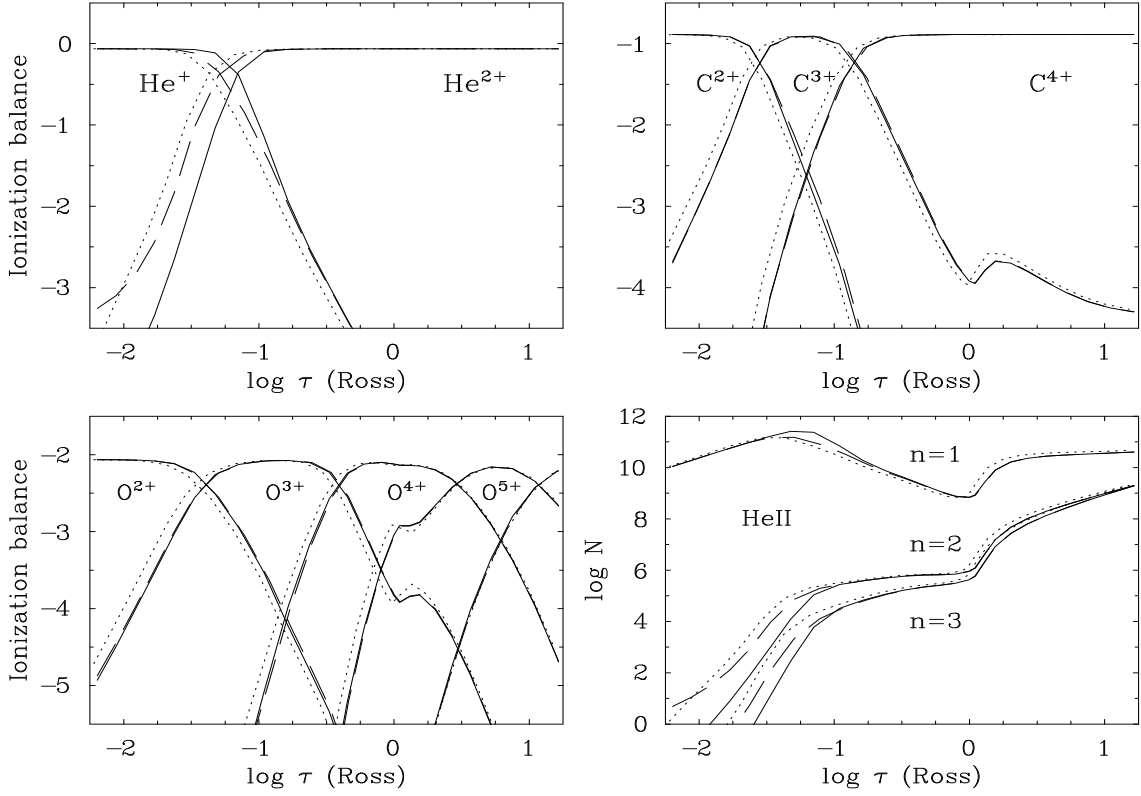}
\caption{Comparison between the ionization structure of helium,
carbon, oxygen and the level population of \heii\ for the
models presented in Sect.~\ref{sec:res} (solid), the model 
with metal lines in the \heii\ Ly$\alpha$ region removed (dashed)
and the model with the reduced $v_{turb}$ (dotted)}
\label{all_fluxes_ion}
\end{figure*}

The elimination of 21 \fevi\ lines, results in a small increase in the wind 
ionization balance with respect to the model of Sect.\ref{sec:res}. Next we eliminated the 
\oiii\ Bowen lines at 305.72 and 303.65~\AA\ 
($^3P-^3P^o$ and $^3P-^3D^o$) and a further increase in
the ionization balance was observed. 
Next we eliminated 1 \cav\ line, with no change in the
calculated spectrum and finally we removed 63 \ciii\ lines, 
which produced a further small increase in the
ionization balance. In Fig.~\ref{all_fluxes} we compare the synthetic spectrum
from the {\it basic model} (solid line) with the 
spectrum from the {\it no-PL model} (dashed line), while
in Fig.~\ref{all_fluxes_ion} we present the corresponding ionization 
structures. The {\it basic model} presents a less ionized spectrum,
with weaker \heii\ lines and stronger \hei\ lines.
The ionization structure of helium is clearly higher in the model without the
metal lines, although that of carbon and oxygen appear similar, even if
the lines of \civ\ follow the trend of the \heii\ lines.
Eliminating some metal lines does indeed change the wind ionization balance. 
The ionization structure and synthetic spectrum of the {\it basic model}
can be matched of we increase the effective temperature of the
{no-PL model} by about 10\,000~K.

As a further test of the effect of line-line interaction
on the overall wind ionization we calculated another model, with the same 
atomic data as in Sect.\ref{sec:res}, but with the
turbulent velocity of the lines in the \heii\ Ly$\alpha$ ($\pm$300~km~s$^{-1}$)
region 
set at 5~km~s$^{-1}$\footnote{This yields a Doppler velocity of 13~km~s$^{-1}$.} 
instead of 50~km~s$^{-1}$.
In this way line-line interaction is reduced, since a
smaller turbulent velocity leads to narrower lines, smaller overlap and therefore 
less interaction. {\it We refer to this model as the ``low-vturb model"}

In Fig.~\ref{all_fluxes} we show the synthetic spectrum for the
{\it low-vturb model} (dotted) compared to the {\it basic model}
(solid), and the {\it no-PL model} (dashed), while in Fig.~\ref{all_fluxes_ion}
we compare their respective ionization structures. The ionization 
equilibrium of the {\it low-vturb model} is higher than for the {\it basic model}
but similar to the {\it no-PL model}.
By reducing the line
widths we have reduced the interaction between lines with an overall shift to a
higher wind ionization, similar to the shift produced by the elimination of 
metal lines near \heii\ Ly$\alpha$. 
  
The interception of continuum photons by 
metal lines (continuum blanketing) is an important effect and the shift
in ionization observed when eliminating metal lines (solid and dashed lines 
in Figs.~\ref{all_fluxes} and \ref{all_fluxes_ion}) {\it could} also 
be due to the lack of continuum blanketing by the lines we eliminated.
The change in ionization between the {\it basic model}
and the {\it low-vturb model} (solid and dotted lines
in Figs.~\ref{all_fluxes} and \ref{all_fluxes_ion}), on the other hand, is due to
a reduced line-line interaction. The similarity of the ionization shifts in the
two trials (dashed and dotted lines in Figs.~\ref{all_fluxes} 
and \ref{all_fluxes_ion}), indicates that
line-line interaction plays an important role in the change
observed when eliminating metal lines. Weak but numerous metal lines,
with wavelengths near that of strong emission lines are important in the
overall ionization balance of the wind and therefore in the derived parameters.

From earlier trials we can estimate that the difference between the {\it basic 
model} (solid)
and the {\it no-PL} or the {\it low-vturb} models
(dashed and dotted, respectively) in Fig.~\ref{all_fluxes} is equivalent to
about 10~kK in effective temperature, but only about 20\%
in bolometric luminosity (because the radius to which the stellar temperature 
refers also changes).
We can therefore conclude that {\it for this parameter space}
photon-loss from the \heii\ Ly$\alpha$,
does lead to a higher luminosity, although not as high as that determined
via the evolutionary mass-luminosity relation. 
 


\section{Discussion}\label{sec:conc}

In the following sections we present a critical summary of the
work carried out in this paper.

\subsection{Stellar parameters and bolometric luminosity}

We have presented a detailed spectroscopic analysis of the
WR star in the $\gamma$~Velorum WR+O binary system. Using the wavelength-dependent light
ratio and the synthetic O star spectrum from Paper~I we have generated
a corrected WR optical and IR spectrum which we have fitted with a synthetic spectrum
using the \cmf\ non-LTE model atmosphere code. Results are summarised in
Table~\ref{tab:wr11_summary}. Our derived mass-loss is in excellent agreement
with the mass-loss derived by Stevens et al. (1996) from X-ray {\sc asca} observations,
once their value is scaled to the {\sc hipparcos} distance. 
Although the parameters of temperature,
mass-loss and carbon abundance are well constrained, the oxygen 
abundance is assumed. The discrepancy between the \heii\ lines at 
4340, 4859 and 6560~\AA\ and other \heii\ lines remains to
be explained. 

\begin{table*}
\begin{center}
\caption{Summary of the stellar parameters obtained for WR11
using {\sc cmfgen} and {\sc isa}-wind (blanketed with photon loss
activated), compared with those derived by Schaerer et al. (1997)
and De Marco et al. (1999a) with a non-line blanketed
model with no photon loss implemented. The results obtained by Dessart et al.
(2000) for WR135 are also included}
\label{tab:wr11_summary}
\begin{tabular}{cccccc}
\hline
Parameters & {\sc cmfgen} & {\sc isa}-wind & De Marco et al. & Schaerer et al. & WR135: Dessart et al. \\
Filling factor, $f$ &  10\% &  100\%  & 100\% & 100\% & 10\% \\
Photon Loss & -- &  yes &  no & no & -- \\
\hline                                   
$T_{\ast}$ (kK)                & 57.1     & 56.0   & 75.9   & 51.0   & 63.0 \\
log($\mdot/\sqrt{f}$) $M_{\odot}$yr$^{-1}$& --4.53 &--4.48  & --4.33 & --4.2 & --4.9 \\
C/He(by number)                & 0.15     & 0.06  & 0.143  & 0.25   & 0.13 \\
O/He(by number)                & 0.03     & 0.01   & --     & --     & 0.03 \\
R (R$_{\odot}$)                & 3.2      & 3.2   & 1.84   & 3.3    & 3.4 \\
log($L_{\ast}/L_{\odot}$)      & 5.00     & 4.95   & 5.01   &  4.8   & 5.20 \\ 
BC (mag)                       & --4.00   & --3.89 & --4.1  & --3.5  & --4.0 \\ 
$v_\infty$ (km~s$^{-1}$)       & 1550     & 1550   & 1300   & 1450   & 1400 \\            
$\eta$/$\sqrt{f}$              & 21       & 28     & 29     &  144   & 5.4   \\
\hline
\multicolumn{4}{l}{$^*$For $\Delta M_V$ = 1.36.}\\
\end{tabular}
\label{tab:wr11_summary}
\end{center}
\end{table*}

Previous analyses of WR11 were carried out by Schaerer et al. (1997) and
by De Marco et al. (1999a - using the stellar atmosphere code of
Koesterke \& Hamann, 1995). Schaerer et al. determined  
the stellar effective temperature by fitting the \hei\ \la 4471 / 
\heii\ \la 4541 ratio, assuming a (higher) mass-loss and
carbon abundance. De Marco
et al. (1999a) derived larger mass-loss and temperature,
and a lower wind terminal velocity. This can be attributed to the fact that
in the parameter space of $\gamma$~Velorum 
the helium spectrum can be reproduced {\it approximately}
with a range of temperature and 
mass-losses. On the other hand, when the lines are analysed in detail, it is
clear that for temperatures larger than about 57~kK and 
log($\dot{M}$/M$_\odot$~yr$^{-1}$)$>$--4.45, 
predicted line shapes deteriorate, becoming wider. This effect
was partly compensated for by De Marco et al. (1999a) by adopting a lower wind
velocity, although it is clear from their fits that most lines
are still too wide. Further, their neglect of line-blanketing and therefore of
photon loss (no line-line interaction between the key lines), 
might have contributed to the different parameters.

With respect to the analysis of Schaerer et al. (1997),
the star becomes twice as luminous, 10\% hotter 
and shows a mass-loss lower by a factor of two. The carbon
abundance is lower by 40\% (although we remind the reader that
Schaerer et al. {\it assumed} both the mass-loss and the carbon abundance). 
This carries implications for studies of wind-wind interaction. From this study and
the results presented in Paper~I for the O star, we determine:

\[
{(\dot{M}\ v_\infty)_{WR} \over (\dot{M}\ v_\infty)_O} \simeq
33
\]

\noindent
compared to a value of 208 derived by 
Schaerer et al. (1997). Adopting the relationships
of Eichler \& Usov (1993) we determine that the region of stellar wind 
collision is approximately
9-30~R$_\odot$ from the O star (0.15 of the distance that separates the two
stars (63-200~R$_\odot$ - Schmutz et al. 1997) vs. 0.06 for the study 
of Schaerer et al. (1997)).
Overall the stellar parameters derived 
are not dissimilar from those derived for the single WC8 star WR135 by
Dessart et al. (2000; see Table~\ref{tab:wr11_summary}). This indicates that
it is unlikely that substantial levels of emission are produced
by the collision region or that an anomalous ionization is induced by the
O star ionizing flux.

The derived spectroscopic luminosity ($\log L/L_{\odot} = 5.0\pm0.1$)
is increased with respect to that found in previous studies, 
         partly due to line-line interaction, and is close to the 
         luminosity derived in Paper~I via the mass-luminosity relationship 
         for WR stars ($\log L/L_{\odot} = 5.2\pm0.1$). An important
         implication of the higher luminosity and of a 10\% clumping factor,
         is that it helps to reduce the
         performance number by a factor of $\sim$20, from 144 to 7, assuming
         a 10\% clumping filling factor. Lucy \& Abbott (1993) showed
         that radiation pressure on spectral lines should in principle
         be able to drive winds with performance numbers $\simeq$ 10.
         So, the strongly reduced $\eta$ suggests that line driving is
         the mechanism responsible for the dense wind of Wolf-Rayet component
         in $\gamma$~Velorum.
 
\subsection{\cmf\ vs. \isa}

Parameters derived with the fast Sobolev approximation code
\isa\ are similar to those derived with \cmf , apart from
the derived carbon mass fraction, with \cmf\ resulting in
a carbon abundance which is a factor of two larger than \isa . 
This discrepancy is not totally understood although we should note
that the co-moving frame code of Koesterke \& Hamann (1995)
compares favourably with \cmf\ for cases in common (e.g. for WR135 -
Dessart et al. 2000). 

A secondary difference is in the fluxes at $\lambda$$<$400~\AA .
Crowther et al. (1999) tested the far UV model fluxes of \cmf\ and \isa\ by
modelling the H~{\sc ii} region associated with the WN8 star WR124,
showing that, for that system, the \isa\ flux was too hard. Also in the
current analysis we find that the \isa\ flux is harder for \la $<$400~\AA :
we suspect that this discrepancy is
due to the lack of photon branching in the
Monte Carlo code, which would lead to distributing energetic photons to
longer wavelengths.
Although the ionising properties of the two synthetic atmospheres
(the values of $Q_0$ and $Q_1$) are similar, the difference in the extreme UV
can carry implications when model WR fluxes are used to interpret nebular line emission
from extra-galactic stellar populations (Leitherer et al. 1999). 

We stress that the comparison between the two codes is reasonable only if photon loss
is included in \isa\ via the approximation detailed in Sect.\ref{ssec:pholos}. 
If photon loss were not accounted for, the determined \isa\ effective temperature 
would be about 8000~K lower, comparable to the difference found in 
Sect.\ref{sec:pholos}
when reducing the interaction between metal lines and \heii Ly$\alpha$.

\subsection{The photon loss effect}

The photon loss effect demonstrated by Schmutz (1997) as the cause of a
lower wind ionization, was tested here using the \cmf\ 
code which treats line-line interaction as part of the blanketing.
For this parameter space the interaction between the \heii\ Ly$\alpha$ and
nearby \oiii\ and \ciii\ 
lines causes a shift to lower ionization equilibrium,
equivalent to $\sim$10\,000~K or a $\sim$20\%
increase in the luminosity\footnote{The quantification of the luminosity shift
resulting from a different temperature depends on the location of the
$R_\ast$, since this can change from model to model, it is not possible to 
simply
derive the change in luminosity from the change in temperature and the Boltzman 
relationship.}.
By reducing the turbulent velocity from 50 to 5~km~s$^{-1}$ for lines residing 
near 303~\AA , we have shown
that the wind ionization is enhanced in a similar way, demonstrating that
line-line interaction, and not continuum blanketing, is at the origin of the ionization shift observed
when eliminating metal lines near the \heii\ Ly$\alpha$.
If line-line interaction between strong resonance lines and weaker but more numerous
metal lines leads to a shift in the wind ionization, a detailed
treatment of even the weakest lines is important.
We remind the reader that our test remains incomplete because, although \cmf\ treats
line-line interaction not all lines are accounted for. Amongst the
ones that are {\it not} treated there might be some that could
play an important 
role in this mechanism. Additionally, our conclusion is only valid within the
parameter space appropriate for WR11, a WC8 star, 
since for different temperature and mass-loss
the strength of the \heii\ and other strong lines will be different
(cf. Crowther et al. (1999) who found that photon-loss did not play a role in
the atmosphere of their WN8 star).

One should conclude from the presence of the photon loss mechanism 
that a proper treatment of the metal lines not only requires one to include
very many lines (to realistically model the blanketing), but also
the very critical ones, such as those near the strong \heii\ Ly$\alpha$
and possibly at other resonance lines such as \civ\ at 1550~\AA . 
Furthermore, the presence 
of photon loss shows that applying mean blanketing factors (Pauldrach 
et al. 1996) - averaged over 20--50~\AA\ intervals - may miss out
some interaction with an overall effect on the derived stellar parameters
for WR stars.
Finally, it is clearly pointed out that macro-turbulence is 
connected with photon loss, as the amount of turbulence essentially dictates the
size and strength of the set of lines involved in the line-line 
interactions.

\acknowledgements
OD acknowledges support from PPARC grant PPA/G/S/1997/00780. PAC 
is funded by a Royal Society University Research Fellowship. 
DJH acknowledges support from NASA grants NAG5-8211 and NAGW-3828.
AdK acknowledges support from NWO Pionier grant 600-78-333 to
L.B.F.M. Waters and from NWO Spinoza grant 08-0 to E.P.J. van
den Heuvel. JS acknowledges the grant Wo 296/22-1,3 by
the Deutsche Forschungsgemeinschaft.
Thanks to Patrick Morris and Karel van der Hucht for
providing the ISO data, which is an ESA project with instruments 
funded by ESA member states (especially the PI countries: France, Germany,
The Netherlands and the United Kingdom) with the partecipation of ISAS
and NASA.

\end{document}